# Human Activity Recognition using Deep Learning Models on Smartphones and Smartwatches Sensor Data

Bolu Oluwalade[1], Sunil Neela[1], Judy Wawira[2], Tobiloba Adejumo[3] and Saptarshi Purkayastha[1]
[1]*Department of BioHealth Informatics, Indiana University-Purdue University Indianapolis, U.S.A.*
[2]*Department of Radiology, Imaging Sciences, Emory University, U.S.A.*
[3]*Federal University of Agriculture, Abeokuta, Nigeria*

Abstract: In recent years, human activity recognition has garnered considerable attention both in industrial and academic research because of the wide deployment of sensors, such as accelerometers and gyroscopes, in products such as smartphones and smartwatches. Activity recognition is currently applied in various fields where valuable information about an individual's functional ability and lifestyle is needed. In this study, we used the popular WISDM dataset for activity recognition. Using multivariate analysis of covariance (MANCOVA), we established a statistically significant difference ($p < 0.05$) between the data generated from the sensors embedded in smartphones and smartwatches. By doing this, we show that smartphones and smartwatches don't capture data in the same way due to the location where they are worn. We deployed several neural network architectures to classify 15 different hand and non-hand oriented activities. These models include Long short-term memory (LSTM), Bi-directional Long short-term memory (BiLSTM), Convolutional Neural Network (CNN), and Convolutional LSTM (ConvLSTM). The developed models performed best with watch accelerometer data. Also, we saw that the classification precision obtained with the convolutional input classifiers (CNN and ConvLSTM) was higher than the end-to-end LSTM classifier in 12 of the 15 activities. Additionally, the CNN model for the watch accelerometer was better able to classify non-hand oriented activities when compared to hand-oriented activities.

## 1 INTRODUCTION

Over the past decade, smartphones have become an important and indispensable aspect of human lives. More recently, smartwatches have also been widely accepted as an alternative to conventional watches, which is referred to as the *quantified self* movement (Swan, 2013). Even in low- and middle-income countries, smartphones have been embedded in the social fabric (Purkayastha et al., 2013), even though smartwatches haven't. Thus, working on both smartphone and smartwatch sensor is still quite relevant. Smartphones and smartwatches contain sensors such as accelerometer, gyroscope, GPS, and much more. These sensors help capture activities of daily living such as walking, running, sitting, etc. and is one of the main motivations for many to own a smartwatch. Previous studies have shown accelerometer and gyroscope to be very effective in recognition of common human activity (Lockhart et al., 2011). In this study, we classified common human activities from the WISDM dataset through the use of deep learning algorithms.

The WISDM (Wireless Sensor Data Mining) Lab in the Department of Computer and Information Science of Fordham University collected data from the accelerometer and gyroscope sensors in the smartphones and smartwatches of 51 subjects as they performed 18 diverse activities of daily living (Weiss, 2019). The subjects were asked to perform 18 activities for 3 minutes each while keeping a smartphone in the subject's pocket and wearing the smartwatch in the dominant hand. These activities include basic ambulation related activities (e.g., walking, jogging, climbing stairs), hand-based daily activities (e.g., brushing teeth, folding clothes), and various eating activities (eating pasta, eating chips) (Weiss, 2019). The smartphone and smartwatch contain both accelerometer and gyroscope sensors, yielding a total of four sensors. The sensor data was collected at a rate of 20 Hz (i.e., every 50ms). Either one of Samsung Galaxy S5 or Google Nexus 5/5X smartphone running the Android 6.0 was used. The LG G Watch was the smartwatch of choice. A total of 15,630,426 raw measurements were collected.

We answer three main research questions in the analysis of this data. Firstly, are there significant differences in how the two devices capture data, even though they both contain similar accelerometer and gyroscope sensors? Secondly, what is the best method to recognize the activities that were performed by the subjects through this sensor data? Thirdly, by using an identified activity, how accurately can we forecast or simulate the activities? We conducted a MANCOVA analysis, which showed that there is a statistically significant difference between the data obtained from the accelerometer and gyroscope in the smartphone and smartwatch. After separating the smartphone and smartwatch data, we classified the activities using neural network architectures, including Long short-term memory (LSTM), Bi-directional Long short-term memory (BiLSTM), Convolutional Neural Network (CNN), and Convolutional LSTM (ConvLSTM). Finally, we developed a GRU model to forecast the last 30 seconds of the watch accelerometer's raw values and calculated the accuracy metrics of this forecasting with the actual values.

## 2 RELATED WORKS

Due to the increase in the availability of several sensors like accelerometer and gyroscope in various consumer products, including wearables, there has been a rise in the number of research studies on human activity recognition (HAR) using sensor data. In one of the earliest HAR studies, Kwapisz et al. used phone accelerometers to classify 6 human activities, including walking, jogging, climbing the stairs, walking down the stairs, sitting, and standing using machine learning models like logistic regression and multilayer perceptron (Kwapisz et al., 2011). Their models recognized most of the activities with an accuracy of 90%. Esfahni et al. created the PAMS dataset containing both smartphone's gyroscope and accelerometer data (Esfahani and Malazi, 2017). Using the section of the data collected from holding the phone with non-dominant hand, they developed multiple machine learning models to identify the same six activities as Kwapisz et al., and obtained a precision of more than 96% for all the models. Random forest and multilayer perceptron models outperformed the rest, with a precision of 99.48% and 99.62% respectively. These results were better than the ones obtained from data collected when the phone was held in the dominant thigh (Esfahani and Malazi, 2017). Also, Schalk et al. obtained more than 94% accuracy for same activities as above by developing a LSTM RNN model (Pienaar and Malekian, 2019). Agarwal et al. proposed a LSTM-CNN Architecture for Human Activity Recognition learning model for HAR. This model was developed by combining a shallow RNN and LSTM algorithm, and its overall accuracy on the WISDM dataset achieved 95.78% accuracy (Agarwal and Alam, 2020). In addition, previous studies like Walse et al. (Walse et al., 2016) and Khin (Oo, 2019) have also used the WISDM accelerometer data to classify a maximum of 6 activities in their work. Although the above models could generally recognize human activities, they were evaluated on their ability to recognize just six human activities and therefore do not provide generalization. Our study addresses these shortcomings by developing several deep learning algorithms for 15 human activities recorded in the WISDM data. We select the best model based on the F1 score, i.e., considering both precision and recall. Here, we obtained an average classification accuracy of more than 91% in our best performing model. Additionally, we try to simulate the data 30 seconds into the future and provide metrics that might be used by other researchers to build more generalizable models.

## 3 METHODOLOGY

### 3.1 WISDM Dataset Description

The WISDM dataset contains raw time series data from phone and watch's accelerometer and gyroscope (X-, Y-, and Z-axis). The raw accelerometer's and gyroscope's signals consist of a value related to each of the three axes. The raw data was segmented into 10-second data without overlapping. Three minutes (180 Seconds) of raw data was divided into eighteen, 10-seconds segment where each segment's range was calculated to obtain 18 values. This was further divided into 10 equal-sized bins to give X 0-9; Y 0-9, Z0-9. Features were generated based on these 10 bins obtained the raw accelerometer and gyroscope readings. Binned distribution, average, standard deviation, variance, average absolute difference, and time between the peaks for each axis were calculated. Apart from these, other features, including Mel-frequency cepstral coefficients, cosine distance, correlation, and average resultant acceleration, were calculated but are not used in this study. After this preprocessing, we finally had the following entries for each of the device sensors:

- phone accelerometer: 23,173

- phone gyroscope: 17,380
- watch accelerometer: 18,310
- watch gyroscope: 16,632

Also, the activities are divided into two classes: non-hand and hand-oriented activities.

- <u>Non-hand</u>-oriented <u>activities</u>: walking, jogging, stairs, standing, kicking, and sitting
- <u>Hand-oriented</u> activities: dribbling, playing catch, typing, writing, clapping, brushing teeth, folding clothes, drinking and eating

## 3.2 MANCOVA Analysis

Multivariate Analysis of (Co)Variance (MANCOVA) explores the relationship between multiple dependent variables, and one or more categorical and/or continuous outcome variables. To perform the MANCOVA analysis on our data, we used the X-, Y-, Z-axis data of the accelerometer and gyroscope at each time epoch as the dependent variables. The categorical variables were the phone and watch. We defined our null hypothesis that there is no statistically significant difference between the phone and watch data. The alternate hypothesis was that there is a statistically significant difference between the phone and watch data. The null hypothesis was rejected as we obtained a statistically significant difference between the phone and the watch data ($p$ less than 0.05).

## 3.3 Classification Models

The MANCOVA analysis showed a significant difference between the phone and watch data. Therefore, we created separate classification models for the phone and watch. We selected 44 features from the watch and phone data with an activity label attached to each row in the dataset. The activities include walking, jogging, walking up the stairs, sitting, standing, typing, brushing teeth, eating soup, eating chips, eating pasta, drinking, eating sandwiches, kicking, playing catch, dribbling a ball, writing, and clapping. Thus, a total of 18 activities. We combined all the different eating activities to form a combined "eating" activity. This reduced the number of activities to 15. We used the Keras package in Python for the experiments. The architecture of the watch accelerometer classification models is described below. In each case of training the models, we stopped the epochs when the training loss became equal to the validation loss to prevent overfitting:

### 3.3.1 Long Short-term Memory (LSTM) Networks

Hochreiter and Schmidhuber originally introduced LSTMs (Hochreiter and Schmidhuber, 1997), and were refined and popularized later (Sherstinsky, 2018) (Zhou et al., 2016). LSTMs are a special kind of recurrent neural networks (RNNs) capable of learning long-term dependencies. This quality in the network architecture helps to remember certain useful parts of the sequence and helps in learning parameters more efficiently. The scaled dataset was input into the LSTM model containing 128 LSTM units, followed by a dropout layer (0.3 units), dense layer (64 Units), dropout layer (0.2 units), dense layer (64 units), dense layer (32 units) and a last dense layer with 15 units (for the number of classes). We used *Softmax* as the activation function in the last layer, *ReLU* for the previous layers, and the *Adam* optimizer. The loss was calculated in *categorical cross-entropy*. The model was trained for 226 epochs with a batch size of 32.

### 3.3.2 Bi-directional Long Short-term Memory (BILSTM) Networks

We also implemented a BiLSTM model to observe the effects of either direction on performance. In the BiLSTM, we fed the data once from the beginning to the end and once from the end to the beginning. By using two hidden states in BiLSTM, we can preserve information from both the past and the future at any point in time. The parameters of our BiLSTM model is similar to the LSTM model.

### 3.3.3 Convolutional LSTM

Xingjian introduced convolutional LSTM's (Shi et al., 2015) in 2015. Convolutional LSTMs (ConvLSTM) are created by extending the fully connected LSTM to have a convolutional structure in both the input-to-state and state-to-state transitions. ConvLSTM network captures spatiotemporal correlations better and outperforms Fully Connected LSTM networks.

The scaled data is reshaped and inputted to a 1-dimensional convolution layer with 128 filters of kernel size 4 followed by a dropout layer (0.4), LSTM layer with 128 units, Dense layer with 100 units, Dense layer with 64 units, Dropout layer with 0.2 dropout rate, Dense layer with 32 units, and finally a Dense layer with 15 units for classification. We used *Softmax* as the activation function in the last layer, *ReLU* for the previous layers, and *Adam* optimizer. The loss was calculated in *categorical cross-entropy*.

The model was trained for 95 epochs with a batch size of 32.

### 3.3.4 Convolutional Neural Network (CNN)

The data is reshaped and inputted to a 1-dimensional convolution layer with 128 filters of kernel size 10 followed by a dropout layer (0.4), 1-dimensional convolution layer with 128 units and 10 kernel size, dropout layer with 0.2 dropout rate, 1-dimensional max-pooling layer with 0.2 pool size, flatten layer, dense layer with 64 units and finally a Dense layer with 15 units for classification. We used *Softmax* as the activation function in the last layer, *ReLU* for the previous layers, and *Adam* optimizer. The loss was calculated in *categorical cross-entropy*. The model was trained for 148 epochs with a batch size of 32.

## 4 RESULTS

### 4.1 Classification Results

To classify the activities, we utilized four classifiers, namely *Long short-term memory (LSTM), Bi-directional Long short-term memory (LSTM), Convolutional Neural Network (CNN), and Convolutional LSTM (ConvLSTM)*. The Precision and F1 scores were used as evaluation metrics to analyze the performance of the classifiers.

Table 1: The Macro-F1 values of different classifiers for **watch** sensor data.

| Models | Accelerometer | Gyroscope | Both |
|---|---|---|---|
| CNN | **0.849** | **0.687** | **0.774** |
| BiLSTM | 0.848 | 0.617 | 0.721 |
| ConvLSTM | 0.843 | 0.658 | 0.754 |
| LSTM | 0.825 | 0.627 | 0.743 |

Table 2: The Macro-F1 values of different classifiers for **phone** sensor data.

| Models | Accelerometer | Gyroscope | Both |
|---|---|---|---|
| CNN | 0.796 | 0.387 | 0.631 |
| BiLSTM | 0.773 | 0.429 | 0.611 |
| ConvLSTM | **0.814** | **0.432** | **0.638** |
| LSTM | 0.756 | 0.395 | 0.743 |

Tables 1 and 2 demonstrate the Macro-F1 values of the different classifiers for the watch and phone data. From the above tables, the classifiers performed better with the watch data. Also, the classifier performed better with accelerometer than gyroscope data. Thus, the classifiers performed best on the watch accelerometer data with a Macro-F1 measure of 0.849, 0.848 and 0.843 and 0.825 for the CNN, BiLSTM, ConvLSTM, and LSTM models, respectively. These classifiers' performance was not greatly different from each other, with all the classifiers having a Macro-F1 measure of more than 0.80 (80%) with the CNN marginally performing the best with a Macro-F1 measure of 0.849 (84.9%).

The watch accelerometer data were divided into training, testing, and validation data, respectively, in 0.8, 0.1, and 0.1 ratios. These data contain 14648, 1831, and 1831 records for training, testing, and validation, respectively. The classifiers' precision values on the watch accelerometer data separated into hand oriented and non-hand oriented classes are presented in Tables 3 and 4 below.

Table 3: The precision values of different classifiers for watch accelerometer data (Non hand-oriented activities).

| Activities | CNN | BiLSTM | ConvLSTM | LSTM | Mean |
|---|---|---|---|---|---|
| Walking | 0.861 | 0.869 | 0.925 | 0.846 | 0.875 |
| Jogging | 0.833 | 0.859 | 0.865 | 0.872 | 0.857 |
| Stairs | 0.909 | 0.939 | 0.900 | 0.862 | 0.903 |
| Sitting | 0.918 | 0.836 | 0.844 | 0.851 | 0.862 |
| Standing | 0.917 | 0.818 | 0.669 | 0.760 | 0.791 |
| Kicking | 0.926 | 0.857 | 0.867 | 0.857 | 0.877 |
| Mean | 0.894 | 0.863 | 0.845 | 0.841 | 0.861 |

Table 4: The precision values of different classifiers for watch accelerometer data (Hand-oriented activities).

| Activities | CNN | BiLSTM | ConvLSTM | LSTM | Mean |
|---|---|---|---|---|---|
| Typing | 0.862 | 0.773 | 0.838 | 0.671 | 0.786 |
| Brushing | **0.981** | **0.955** | **0.963** | **0.972** | **0.968** |
| Drinking | 0.867 | 0.808 | 0.801 | 0.768 | 0.811 |
| Eating | 0.757 | 0.884 | 0.677 | 0.779 | 0.774 |
| Catch | 0.850 | 0.871 | 0.870 | 0.906 | 0.874 |
| Dribbling | 0.844 | 0.802 | 0.797 | 0.768 | 0.802 |
| Writing | 0.807 | 0.816 | 0.906 | 0.836 | 0.841 |
| Clapping | 0.785 | 0.902 | 0.758 | 0.769 | 0.804 |
| Folding | 0.821 | 0.830 | 0.871 | 0.794 | 0.829 |
| Mean | 0.842 | 0.849 | 0.831 | 0.807 | 0.832 |

Tables 3 and 4 show that the precision of the convolutional input classifiers (CNN and ConvLSTM) was higher than the end-to-end LSTM classifiers. The convolutional classifiers gave the best classifications in 12 of 15 activities irrespective of the class of the activities (i.e., non-hand-oriented and hand-oriented activities). The convolution input layer has also been shown to outperform conventional fully connected LSTM in capturing spatiotemporal correlations in data (Shi et al., 2015). Another fact that can be inferred from the table 3 above is that the classifiers perform better in the non-hand-oriented activities like standing, stairs, etc. than the hand-oriented activities like eating, clapping, etc. with the exception of brushing teeth, which has the highest precision values of all the classifiers.

## 4.2 Forecasting Results

In addition to the classification models, we also predicted the last 30 seconds of the raw data for all the eating activities. To achieve this, we utilized a gated recurrent unit (GRU) model. We used the Root Mean Square Error (RMSE), Mean Square Error, Mean Absolute Percentage Error (MAPE), and symmetric Mean Absolute Percentage Error (sMAPE) as evaluation metrics.

Table 5: The different performance metrics of different activities for phone data.

| Activities | RMSE | MSE | MAPE | sMAPE |
|---|---|---|---|---|
| H (eating soup) | 0.078 | 0.00603 | 126.580 | 40.57 |
| I (eating chips) | 0.070 | 0.00494 | 11.43 | 10.59 |
| J (eating pasta) | 0.039 | 0.00152 | 7.263 | 7.25 |
| K (drinking from cup) | 0.083 | 0.00687 | 13.080 | 13.296 |
| L (eating sandwich) | 0.050 | 0.00247 | 4.45 | 4.54 |
| Mean | 0.064 | 0.0044 | 32.56 | 15.25 |

Table 5 shows that GRU gave the best forecast for eating sandwiches when compared to other activities like drinking from a cup, eating pasta, eating chips, eating soup, etc. Also, it can be inferred from Table 5 that the MAPE overstated the error found in activity H because of the presence of outliers when compared to the values of other activities with their respective sMAPE.

## 5 DISCUSSION

The statistically significant difference ($p < 0.05$) between the same kind of sensors in smartphones and smartwatches using a MANCOVA analysis points to some interesting observations. This is likely due to the location of the pocket in comparison to the hand. However, there is more to it than just the differences in the X-, Y-, and Z-axis values due to the height of the sensors from the ground. Our analysis showed that the difference between the peaks and the throughs was also larger in the smartwatch. Furthermore, identifying a constant that differentiated the X-, Y-, or Z-axis between the two devices was practically impossible. This means that the difference between the sensors is not purely due to the height from the ground. Following this distinction, we created various deep learning models to classify 15 different activities based on the smartwatch accelerometer data only. Our findings suggest that non-hand-oriented activities like standing, stairs, etc. are better generalized and better classified than hand-oriented activities like eating, clapping, etc. with the exception of brushing teeth as the classifiers performed better and had higher precision values. This implies that the smartwatch accelerometer data can better classify non-hand oriented activities even if the smartwatch is located on the dominant hand. A recent study by Agarwal et al. (Agarwal and Alam, 2020), used a lightweight RNN-LSTM architecture to classify 6 different non-hand oriented activities based on smartphone accelerometer data. They obtained an average of 0.9581 (95.81%) precision. Our best model obtained a precision to 92.6% for kicking and 98.1% for brushing teeth. We obtained an average of 89% precision for 6 non-hand oriented activities recognition and 84% for 9 hand oriented activities. However, these are not directly comparable, as we also calculated the Macro-F1 values, which consider both the precision and recall i.e., when our model accurately classifies the activity but also fails at classifying the activity. This should be considered as a better measure for HAR. Moreover, we also did better at HAR compared to other papers that used the WISDM data and used LSTM-CNN, and shallow RNN. Accurate HAR is clinically relevant, not only because individuals have started using wearables widely, but also because there are many clinically-relevant sensors such as the BioStamp MC10, fall detection devices, and telemedicine sensors. These sensors might also be useful to identify mental health issues (Addepally and Purkayastha, 2017) or motor function disorders (Ellis et al., 2015) using mHealth apps. All these devices have accelerometer and gyroscope sensors to understand patient's gait, posture, and stability for accurate measurement of other clinical features. As home healthcare, senior home care and health care outside the hospital settings become more common, the application of HAR is becoming more relevant.

We also created a GRU model to forecast the last 30 seconds of raw data generated by the watch accelerometer for 4 eating activities based on the previous 210 seconds data. We obtained an average RMSE of 0.064, which implies a minimal difference from the actual values. Thus, we can say that generating such sensor data for generalizing models might also be a feasible approach for retraining models or transfer learning of models to similar sensors of other devices. This is a future direction that we are pursuing.

# 6 LIMITATIONS

Although the use of classification models such as Long short-term memory (LSTM), Bi-directional Long short-term memory, Convolutional Neural Network (CNN) and Convolutional LSTM, etc., is a fairly common approach to predicting the movement of a person, this study does not provide the needed generalization with the hand-oriented activities. We have not, for example, examined differences in the performance metrics of the eating activity when forecasting the raw values of the last 30 seconds of the watch accelerometer.

# 7 CONCLUSION

In this study, we classified smartphone and smartwatch accelerometer and gyroscope data. We classified the majority of the activities using artificial neural network algorithms, including Long short-term memory (LSTM), Bi-directional Long short-term memory, Convolutional Neural Network (CNN), and Convolutional LSTM. Our classification analysis on 15 different activities resulted in an average classification accuracy of more than 91% in our best performing model. Although previous findings indicated that 6 human activities were used during the analysis, our study followed several 15 human activities, which are better generalized than those in major studies conducted previously. It is possible that outcomes would vary if over 20 or 25 human activities are used. Future researchers should consider investigating the impact of more human activities. Nonetheless, our results provide the needed generalization for non-hand oriented activities recognition cases only.

# REFERENCES

Addepally, S. A. and Purkayastha, S. (2017). Mobile-application based cognitive behavior therapy (cbt) for identifying and managing depression and anxiety. In *International Conference on Digital Human Modeling and Applications in Health, Safety, Ergonomics and Risk Management*, pages 3–12. Springer.

Agarwal, P. and Alam, M. (2020). A lightweight deep learning model for human activity recognition on edge devices. *Procedia Computer Science*, 167:2364–2373.

Ellis, R. J., Ng, Y. S., Zhu, S., Tan, D. M., Anderson, B., Schlaug, G., and Wang, Y. (2015). A validated smartphone-based assessment of gait and gait variability in parkinson's disease. *PLoS one*, 10(10):e0141694.

Esfahani, P. and Malazi, H. T. (2017). Pams: A new position-aware multi-sensor dataset for human activity recognition using smartphones. In *2017 19th International Symposium on Computer Architecture and Digital Systems (CADS)*, pages 1–7. IEEE.

Hochreiter, S. and Schmidhuber, J. (1997). Long short-term memory. *Neural Comput.*, 9(8):1735–1780.

Kwapisz, J. R., Weiss, G. M., and Moore, S. A. (2011). Activity recognition using cell phone accelerometers. *ACM SigKDD Explorations Newsletter*, 12(2):74–82.

Lockhart, J. W., Weiss, G. M., Xue, J. C., Gallagher, S. T., Grosner, A. B., and Pulickal, T. T. (2011). Design considerations for the wisdm smart phone-based sensor mining architecture. In *Proceedings of the Fifth International Workshop on Knowledge Discovery from Sensor Data*, pages 25–33.

Oo, K. K. (2019). Daily human activity recognition using adaboost classifiers on wisdm dataset.

Pienaar, S. W. and Malekian, R. (2019). Human activity recognition using lstm-rnn deep neural network architecture. In *2019 IEEE 2nd Wireless Africa Conference (WAC)*, pages 1–5. IEEE.

Purkayastha, S., Manda, T. D., and Sanner, T. A. (2013). A post-development perspective on mhealth–an implementation initiative in malawi. In *2013 46th Hawaii International Conference on System Sciences*, pages 4217–4225. IEEE.

Sherstinsky, A. (2018). Fundamentals of recurrent neural network (RNN) and long short-term memory (LSTM) network. *CoRR*, abs/1808.03314.

Shi, X., Chen, Z., Wang, H., Yeung, D., Wong, W., and Woo, W. (2015). Convolutional LSTM network: A machine learning approach for precipitation nowcasting. *CoRR*, abs/1506.04214.

Swan, M. (2013). The quantified self: Fundamental disruption in big data science and biological discovery. *Big data*, 1(2):85–99.

Walse, K., Dharaskar, R., and Thakare, V. (2016). Performance evaluation of classifiers on wisdm dataset for human activity recognition. In *Proceedings of the Second International Conference on Information and Communication Technology for Competitive Strategies*, pages 1–7.

Weiss, G. M. (2019). Wisdm smartphone and smartwatch activity and biometrics dataset. *UCI Machine Learning Repository: WISDM Smartphone and Smartwatch Activity and Biometrics Dataset Data Set*.

Zhou, P., Qi, Z., Zheng, S., Xu, J., Bao, H., and Xu, B. (2016). Text classification improved by integrating bidirectional LSTM with two-dimensional max pooling. In *Proceedings of COLING 2016, the 26th International Conference on Computational Linguistics: Technical Papers*, pages 3485–3495, Osaka, Japan. The COLING 2016 Organizing Committee.